\documentclass[conference]{IEEEtran}
\IEEEoverridecommandlockouts

\usepackage{amsmath,amssymb,amsfonts}
\usepackage{graphicx}
\usepackage{textcomp}
\usepackage{xcolor}
\usepackage{cite}
\usepackage{booktabs}
\usepackage[caption=false,font=footnotesize]{subfig} % IEEEtran-fr
\usepackage{placeins}   
\usepackage{url}

\begin{document}

\title{Leveraging Orbital Dynamics with RF Signal Features for Satellite Multi-Orbit \\Proximity Threat Detection}

\author{
\IEEEauthorblockN{Anouar Boumeftah\IEEEauthorrefmark{1}, Gunes Karabulut Kurt\IEEEauthorrefmark{1}}
\IEEEauthorblockA{\IEEEauthorrefmark{1}Poly-Grames Research Center, Department of Electrical Engineering,\\
Polytechnique Montréal, Montréal, QC, Canada \\
\{anouar.boumeftah, gunes.kurt\}@polymtl.ca}
}

\maketitle
\setlength\parindent{0pt}
\begin{abstract}
Proximity-based interference is a growing threat to satellite communications, driven by dense multi-orbit constellations and increasingly agile adversarial maneuvers. We propose a hybrid simulation framework that integrates orbital maneuver modeling with RF signal degradation analysis to detect and classify suspicious proximity operations. Using the open-source Maneuver Detection Data Generation (MaDDG) library from MIT Lincoln Laboratory, we generate labeled datasets combining impulsive maneuver profiles with radio-frequency (RF) impacts across a range of behavioral intents—routine station-keeping, covert shadowing, and overt jamming. Our approach fuses kinematic features such as range, velocity, acceleration, and Time of Closest Approach (TCA), with RF metrics including Received Signal Strength Indicator (RSSI), throughput, and Jammer-to-Signal Ratio (JSR). These features are further enhanced with temporal derivatives and rolling-window statistics to capture subtle or transient interference patterns. A Random Forest classifier trained on this fused feature set achieves 94.67\% accuracy and a macro F1 score of 0.9471, outperforming models using only kinematic or RF inputs. The system is particularly effective in detecting covert threats, such as surveillance or intermittent jamming, that evade RF-only methods.
\end{abstract}


\section{Introduction}
The expansion of multi-orbit satellite constellations---composed of Low Earth Orbit (LEO), Medium Earth Orbit (MEO), and Geostationary Earth Orbit (GEO)---has introduced a new class of threats that combine proximity-based maneuvers with intentional radio-frequency (RF) interference. These threats blur the boundary between orbital dynamics and communication-layer attacks, complicating traditional detection methods that treat these domains in isolation~\cite{leomanni2020, pietro2022}. Satellites capable of dynamic repositioning can exploit relative motion and boresight alignment to degrade uplink or inter-satellite links through targeted jamming, surveillance shadowing, or covert pursuit maneuvers. While the orbital domain provides valuable cues---such as range rate, angular velocity, and maneuver frequency---signal-level observables such as signal-to-jammer-plus-noise ratio (SJNR), received signal strength indicator (RSSI) and throughput offer additional indicators of malicious activity~\cite{yang2024, park2023}.\\

Prior work has shown that abrupt signal degradations often correlate with adversarial kinematics ~\cite{yang2024}, motivating the need for cross-domain frameworks that jointly reason over geometry and RF behavior. Such approaches support early intent inference and more robust threat attribution in contested space environments. This work proposes a hybrid simulation-based framework that fuses orbital maneuver profiles with communication-layer signal degradation to detect and classify proximity-based threats. Using datasets generated with the Maneuver Detection Data Generation (MaDDG) library ~\cite{tierney2025maddg}, we model attacker trajectories across LEO, MEO, and GEO, simulate corresponding RF link degradation, and extract fused features for threat classification. Our approach captures orbit-dependent vulnerability, supports geometry-aware interference modelling, and lays the foundation for explainable and proactive space domain awareness.

\section{Related Work}

Traditional approaches to Space Situational Awareness (SSA) and spectrum monitoring have treated orbital and RF domains as separate problems. In the orbital domain, maneuver detection typically leverages relative orbital element (ROE) analysis, nodal element parameterizations, or angles-only tracking to infer proximity behaviors such as station-keeping, rendezvous, or pursuit~\cite{park2023, roberts2007}. More recent methods employ learning-based sequence models, such as Long Short-Term Memory networks (LSTMs), to distinguish between benign and adversarial maneuvers using temporal dynamics~\cite{riccardo2025, siew2025}. 

In parallel, learning-based RF interference detection methods rely on signal processing techniques like carrier-to-noise ratio (C/N$_0$) monitoring, spectrogram analysis, and automatic gain control tracking~\cite{ghanney2020}. Deep learning pipelines---particularly convolutional autoencoders (CAE) along with object detection algorithms (YOLO-v3)---have demonstrated effectiveness in classifying jamming and spoofing attacks under low signal-to-noise ratio (SNR) conditions~\cite{ghanney2020}. These methods are increasingly of interest to be applied onboard satellites, turning them into distributed interference monitors~\cite{yunfan2024}. Neural network-based Global Navigation Satellite System (GNSS) detectors~\cite{mehr2022} and onboard payloads~\cite{vasseur2021} now autonomously localize and classify uplink interference, while hybrid approaches combine kinematics with RF metrics to improve jamming detection~\cite{boumeftah2025_asms}. Our work builds on this direction by fusing both domains to classify intent and assess vulnerability across orbital layers.\\

We should note that both domains face critical limitations when used independently. Orbital-domain methods may detect close approaches but cannot determine if communication links are affected. RF-only models may flag signal degradation but cannot attribute it to specific geometric configurations or intentional maneuvers. Recent studies have argued for geometry-driven vulnerability assessments, where off-axis distance, relative angular velocity, and line-of-sight alignment serve as critical inputs to interference susceptibility models. Cross-orbit scenarios—such as interference between non-Geostationary orbits (NGSO) and GEO satellites—have highlighted the need for multi-orbit models that incorporate visibility windows and constellation geometry~\cite{yunfan2024}.\\

To overcome these limitations, emerging research has explored multi-modal fusion. Feature-level fusion methods combine orbital kinematics with RF observables like SJNR slope and RSSI variance~\cite{siew2025}. At the model level, Bayesian filters augmented with neural network measurements---such as neural-network-augmented unscented Kalman filters (NN-UKF)---have improved robustness under sparse or noisy data~\cite{park2023}. Decision-level frameworks use explainable AI tools, such as SHAP, to attribute anomalies to either domain. Still, these approaches are hindered by the lack of large-scale, joint-domain datasets. Most public orbital datasets rely on TLEs, which lack maneuver resolution~\cite{roberts2007}, while RF benchmarks often derive from ground-based GNSS interference tests that do not reflect on-orbit conditions~\cite{ghanney2020}.\\

Our work addresses these gaps by proposing a unified dataset and classification pipeline for proximity-based interference. Unlike prior efforts focused on either kinematic or signal features, we explicitly couple orbital state vectors with link-layer degradations to infer intent using a unified feature basis. This enables more accurate detection of threats such as coordinated shadowing, deliberate jamming, and cross-orbit interference. This paper presents the following key contributions: 
\begin{itemize}
    \item We introduce a modular pipeline that combines orbital maneuver simulation using MaDDG with RF signal degradation modeling, producing a labeled dataset of multi-orbit threat scenarios. 
    \item We systematically evaluate how orbit type (LEO, MEO, GEO), attacker geometry, and signal path alignment influence degradation severity, offering insights into orbit-specific vulnerabilities. 
    \item We propose a set of statistically derived RF features—including gradients, rolling variances, and anomaly flags—to capture signal instability without relying on direct SJNR input, enhancing model robustness. 
    \item We train and evaluate a Random Forest machine learning model on combined orbital and RF features to detect and classify proximity-based threats with improved fidelity.
    \item We demonstrate that our model maintains performance without access to simulation-privileged features, improving realism and applicability to operational satellite communications (SatCom) systems.
\end{itemize}

The remainder of this paper is organized as follows: Section~\ref{sec:methodology} details the simulation methodology, including orbital maneuver generation, RF signal modeling, and kinematic feature extraction. Section~\ref{sec:results} presents the classification results, evaluating the effectiveness of RF-only, kinematic-only, and fused models, along with ablation studies and noise-variation experiments. Finally, Section~\ref{sec:conclusion} concludes the paper and outlines future directions for autonomous threat detection in contested orbital environments.

\section{Methodology}
\label{sec:methodology}

This section outlines the simulation framework and methodological components used to generate synthetic threat scenarios in the space domain, incorporating both kinematic and RF signal models. The objective is to create a comprehensive dataset that enables the classification of potentially malicious satellite behavior through learned representations of maneuver and signal interference patterns.

\subsection{Scenario Simulation Framework}

We developed a custom simulation environment designed to model interactions between a GEO communications satellite (hereafter referred to as the ``target''), a ground station (GS), and a second, potentially adversarial satellite (hereafter referred to as the ``attacker''). The framework supports multiple classes of attacker behavior and orbital configurations, enabling robust scenario diversity. Three attacker behavior classes were defined: 

\begin{itemize}
    \item \textbf{Benign:} Satellites maintaining non-threatening orbital separations and minimal RF interference activity.
    \item \textbf{Covert:} Satellites exhibiting subtle maneuvering and intermittent RF jamming, indicative of stealth objectives.
    \item \textbf{Threatening:} Satellites demonstrating aggressive proximity maneuvers and sustained interference activity, aligned with jamming or spoofing intent.
\end{itemize}

To ensure representational diversity, attacker satellites were instantiated across three orbital regimes: LEO, MEO, and GEO. Orbit instantiations were randomized within a constrained domain for each regime, and a single impulsive maneuver ($\Delta v$) was applied within a randomized time window for the corresponding behavior class, with magnitude and direction sampled according to class priors. All orbital propagations were performed in the Earth-Centered Inertial (ECI) frame using the SGP4 propagator and subsequently transformed to the Local-Vertical Local-Horizontal (LVLH) frame of the GEO satellite for feature computation.

\FloatBarrier
\begin{figure*}[!t]
    \centering
    \includegraphics[width=0.85\textwidth]{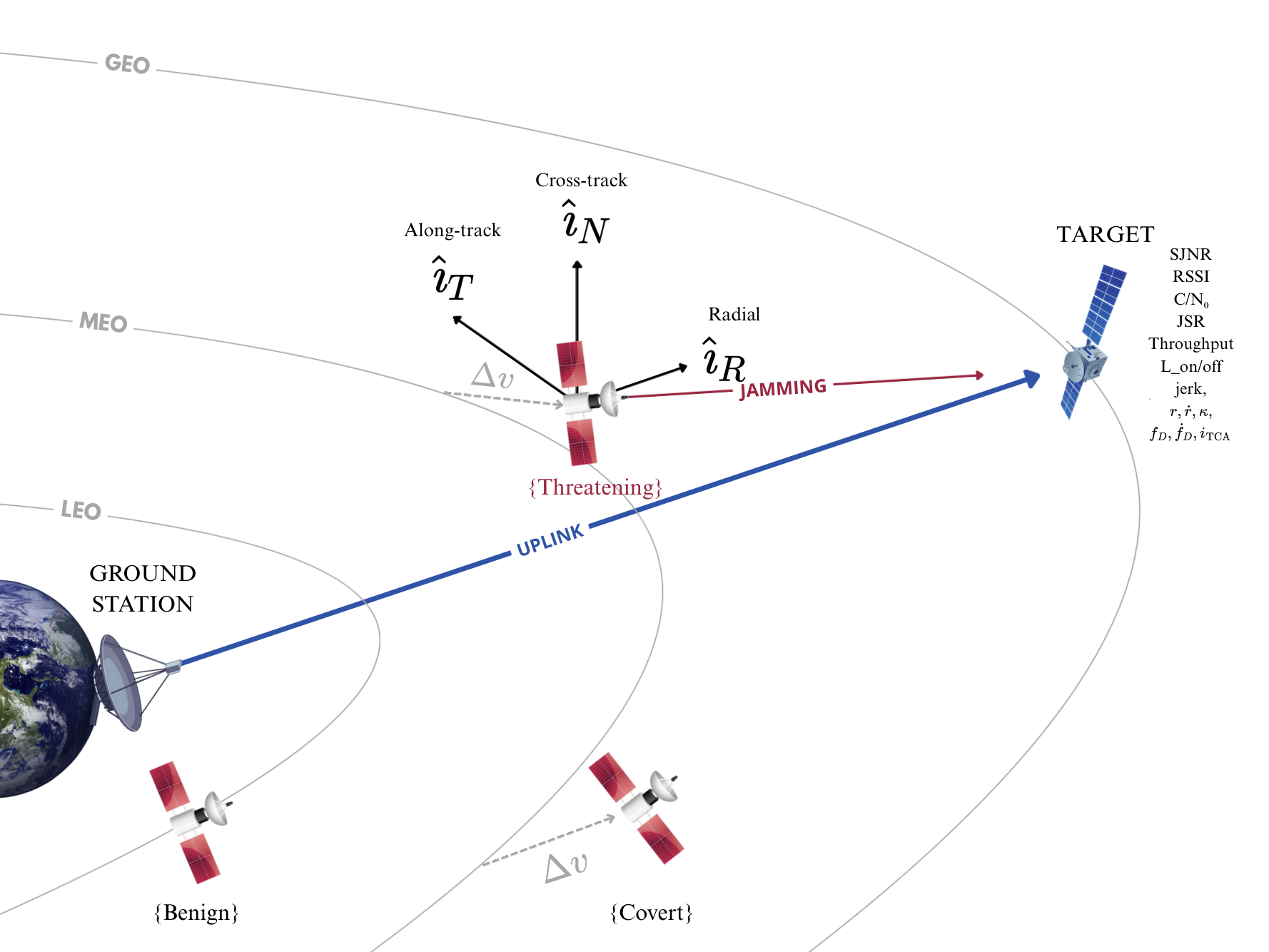}
    \caption{System model illustration: multi-orbit satellite-to-satellite jamming threat scenarios on a GEO target.}
    \label{fig:system_model}
\end{figure*}

\subsection{RF Signal and Interference Modeling}

Each scenario models uplink RF signal links between the ground station and the GEO satellite, alongside a jamming link from the attacker to the GEO receiver. The signal model incorporates antenna gain pattern, pointing jitter, Doppler effects, thermal noise, and RF front-end imperfections. One of the primary metrics for the quality of the downlink is the Signal-to-Jammer Noise Ratio (SJNR), derived from standard link budget equations and jammer incident angle:
\begin{equation}
    \mathrm{SJNR}_{\mathrm{lin}}=\frac{P_{r,\mathrm{sig}}}{P_{r,\mathrm{jam}}+N}, 
    \qquad 
    \mathrm{SJNR}_{\mathrm{dB}}=10\log_{10}\!\left(\mathrm{SJNR}_{\mathrm{lin}}\right),
\end{equation}
where $P_{r,\mathrm{sig}}$ and $P_{r,\mathrm{jam}}$ denote the received signal and jammer power at the GEO receiver, respectively, and $N$ represents the thermal noise power. The Received Signal Strength Indicator (RSSI), incorporating power loss due to free-space propagation and beam misalignment:
\begin{equation}
    \mathrm{RSSI}_{\mathrm{dBm}}=P_{r,\mathrm{sig}}^{\mathrm{dBW}}+30+\varepsilon_{\mathrm{RSSI}},
\end{equation}
where $\varepsilon_{\mathrm{RSSI}}\sim\mathcal{N}\!\left(0,\sigma_{\mathrm{RSSI}}^2\right)$ represents Gaussian measurement noise. The throughput is estimated using the Shannon-Hartley capacity formula under the observed linear SJNR, with additive modeling noise:
\begin{equation}
    \mathrm{Throughput}=\frac{B\,\mathrm{Hz}}{10^{6}}\log_{2}\!\left(1+\mathrm{SJNR}_{\mathrm{lin}}\right),
\end{equation}
where $B$ is the link’s bandwidth.

Link reliability is further characterized via the Carrier-to-Noise Density Ratio (C/N$_0$) and Jammer-to-Signal Ratio (JSR), serving as proxies for link reliability and jamming severity:
\begin{equation}
    \mathrm{C/N}_0~[\mathrm{dBHz}]=P_{r,\mathrm{sig}}^{\mathrm{dBW}}-kB^{\mathrm{dBW}}+10\log_{10}T_{\mathrm{sys}},
\end{equation}
\begin{equation}
    \mathrm{JSR}_{\mathrm{dB}}=P_{r,\mathrm{jam}}^{\mathrm{dBW}}-P_{r,\mathrm{sig}}^{\mathrm{dBW}}.
\end{equation}

The jammer uses a class-conditional burst model, with ON/OFF activity governed by a hidden Markov process. Let $s[i]\in\{0,1\}$ denote the jammer state at time index $i$, where 1 indicates active jamming. The ON and OFF durations are drawn from geometric distributions:
\[
L_{\mathrm{on}}\sim\mathrm{Geom}(p_{\mathrm{on}}), 
\qquad 
L_{\mathrm{off}}\sim\mathrm{Geom}(p_{\mathrm{off}}),
\]
with final segment length clamped between 1 and 500 samples:
\begin{equation}
    L=\min\!\left(\max\{L,\,1\},\,500\right).
\end{equation}

For instance, covert scenarios employ sparse and short-duration jamming bursts, whereas threatening scenarios exhibit frequent, longer-duration interference, this is detailed in Section 2.4. RF perturbations such as Doppler shift, Doppler rate, and Carrier Frequency Offset (CFO) noise are computed at each timestep, with additive Gaussian noise introduced to simulate measurement and hardware uncertainty.

\subsection{Kinematic Feature Computation}

To capture attacker motion signatures, we extract a comprehensive set of kinematic features from the relative dynamics between the attacker and the GEO satellite. 

\subsubsection{Relative Motion Features}

The attacker’s position, velocity, and acceleration vectors are projected into the LVLH coordinate system centered on the GEO satellite. The attacker’s relative position and velocity vectors are first projected into radial (R), along-track (T), and cross-track (N) components. The orthonormal basis is defined as reminded in \cite{casotto2016}: 
\begin{align}
\hat{\mathbf{R}} &= \frac{\mathbf{r}_{\mathrm{tgt}}}{\|\mathbf{r}_{\mathrm{tgt}}\|}, \\
\mathbf{T}' &= \mathbf{v}_{\mathrm{tgt}} - (\mathbf{v}_{\mathrm{tgt}} \!\cdot\! \hat{\mathbf{R}})\, \hat{\mathbf{R}}, \\
\hat{\mathbf{T}} &= \frac{\mathbf{T}'}{\|\mathbf{T}'\|}, \\
\hat{\mathbf{N}} &= \hat{\mathbf{R}} \times \hat{\mathbf{T}}.
\end{align}

Let $\mathbf{r}_{\mathrm{rel}}=\mathbf{r}_{\mathrm{att}}-\mathbf{r}_{\mathrm{tgt}}$, and $\mathbf{v}_{\mathrm{rel}}=\mathbf{v}_{\mathrm{att}}-\mathbf{v}_{\mathrm{tgt}}$. The scalar range and range rate are computed as:
\begin{equation}
r=\|\mathbf{r}_{\mathrm{rel}}\|,\qquad 
\dot{r}=\frac{\mathbf{v}_{\mathrm{rel}}\cdot \mathbf{r}_{\mathrm{rel}}}{\|\mathbf{r}_{\mathrm{rel}}\|}.
\end{equation}

We compute acceleration components in each axis using central differences with time step $\Delta t$:
\begin{align}
a_R[i]&\approx\frac{v_R[i+1]-v_R[i-1]}{2\Delta t},\\
a_T[i]&\approx\frac{v_T[i+1]-v_T[i-1]}{2\Delta t},\\
a_N[i]&\approx\frac{v_N[i+1]-v_N[i-1]}{2\Delta t}.
\end{align}

Additionally, higher-order derivatives such as jerk and curvature are computed as indicators of non-Keplerian behaviour, jerk is approximated as in \cite{sparavigna2015} and curvature is computed using the Frenet-Serret formulation \cite{gray1993}:
\begin{equation}
\mathrm{jerk}[i]\approx\frac{\|\mathbf{a}\|[i+1]-\|\mathbf{a}\|[i-1]}{2\Delta t},
\end{equation}
\begin{equation}
\kappa_i=\frac{\left\|\mathbf{v}_{\mathrm{LVLH},i}\times\mathbf{a}_{\mathrm{LVLH},i}\right\|}
{\left\|\mathbf{v}_{\mathrm{LVLH},i}\right\|^{3}+\varepsilon},
\qquad \varepsilon=10^{-12}.
\end{equation}

Equation (10) includes a small regularization term $\varepsilon$ in the denominator to avoid cases where the velocity magnitude is near zero. This can occur, for instance, in benign behavior, where the object exhibits little or no motion in the LVLH frame. Without this term, curvature would become numerically unstable or undefined in such scenarios.

\subsubsection{Geometric and Temporal Features}

Additional features include Doppler characteristics of the jammer with respect to the GEO receiver. The Doppler shift and its time derivative are given by:
\begin{equation}
\lambda=\frac{c}{f_c},\ \ 
f_D=-\frac{\dot{r}}{\lambda},\ \ 
\dot{f}_D[i]\approx\frac{f_D[i+1]-f_D[i-1]}{2\Delta t},
\end{equation}
where $\lambda = c / f_c$ is the carrier wavelength. The Time of Closest Approach (TCA) is estimated as described in \cite{hejduk2017} as the time corresponding to minimum range between target and attacker:
\begin{align}
i^*_{\mathrm{TCA}} &= \arg\min_i r[i], \\
t_{\mathrm{TCA\_frac}} &= \frac{i^*_{\mathrm{TCA}}}{N_{\mathrm{TCA}} - 1}, \\
t_{\mathrm{toTCA}}[i] &= \frac{i - i^*_{\mathrm{TCA}}}{N_{\mathrm{TCA}} - 1}.
\end{align}

The angle of incidence between the attacker’s position and the GEO receiver’s antenna boresight is computed to assess the geometric alignment of potential interference. In addition, we extract basic temporal statistics—including the minimum, maximum, mean, and slope—of key kinematic variables over time. To capture localized variations, rolling-window metrics such as a 3-point standard deviation and gradient are applied to RSSI and throughput time series. These localized derivatives emphasize abrupt changes or fluctuations that may not be evident from global trends, providing finer granularity for detecting transient jamming events or dynamic behavior in the RF environment.

\subsection{Feature Agregation and Dataset Construction}
\label{subsec:feature-agg}

We generate 3,600 scenarios spanning benign, covert, and threatening behaviors across LEO, MEO, and GEO, each simulated over a 0.1-day horizon (144 minutes) with 10-second resolution (864 samples/scenario), yielding over 3.1 million samples and 400 scenarios per class across \{benign, covert, threatening\} $\times$ \{LEO, MEO, GEO\}. Thus, the corpus comprises 3,600 scenarios (3,110,400 rows). Per-timestep features capture both orbital dynamics and RF behavior, augmented with short-term temporal statistics and scenario-level aggregates. Jammer activity is modeled with bursty ON/OFF patterns rather than independent toggling to reflect realistic interference as summarized in Table~\ref{tab:jammer_params}.

\begin{table}[h]
\centering
\caption{Class Parameters for jammer activation}
\label{tab:jammer_params}
\begin{tabular}{@{}lccc@{}}
\toprule
\textbf{Class} & $p_{\text{off}}$ & $p_{\text{on}}$ & $P(\text{jam})$ \\
\midrule
Benign      & 0.8  & 0.05 & $\sim$6\%  \\
Covert      & 0.6  & 0.15 & $\sim$20\% \\
Threatening & 0.15 & 0.5  & $\sim$77\% \\
\bottomrule
\end{tabular}
\end{table}

To support analysis, three feature views are constructed: RF-only, kinematics-only, and a fused set that combines kinematic, RF, and interaction terms. Labels are assigned at the scenario level (benign, covert, threatening), with metadata such as orbit, scenario identifier, and visibility preserved for later use.

\section{Results}
\label{sec:results}

\begin{figure*}[!t]
    \centering
    \subfloat[RF-only]{\includegraphics[width=0.33\textwidth]{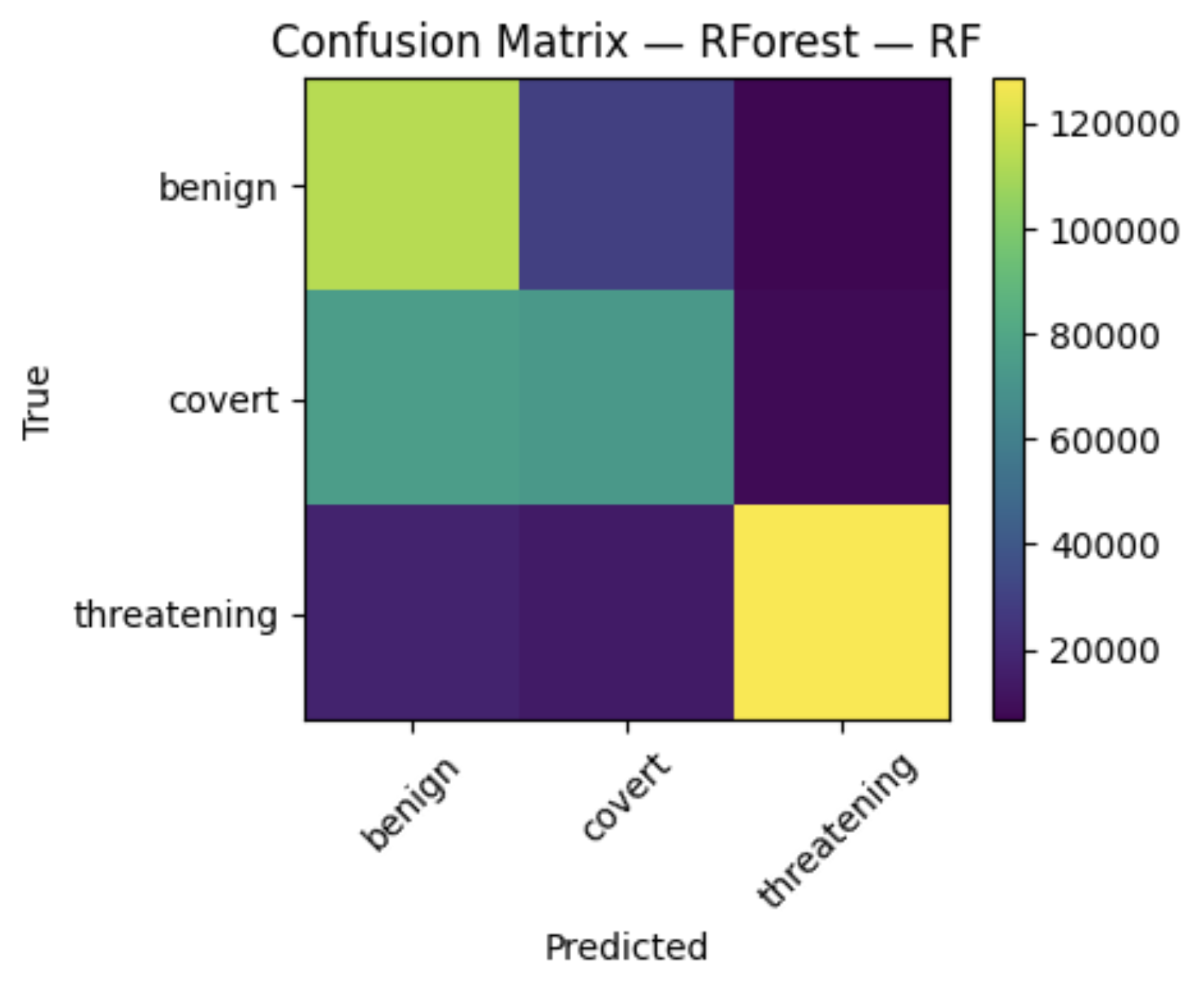}\label{fig:conf_rf}}
    \hfill
    \subfloat[Kinematic-only]{\includegraphics[width=0.33\textwidth]{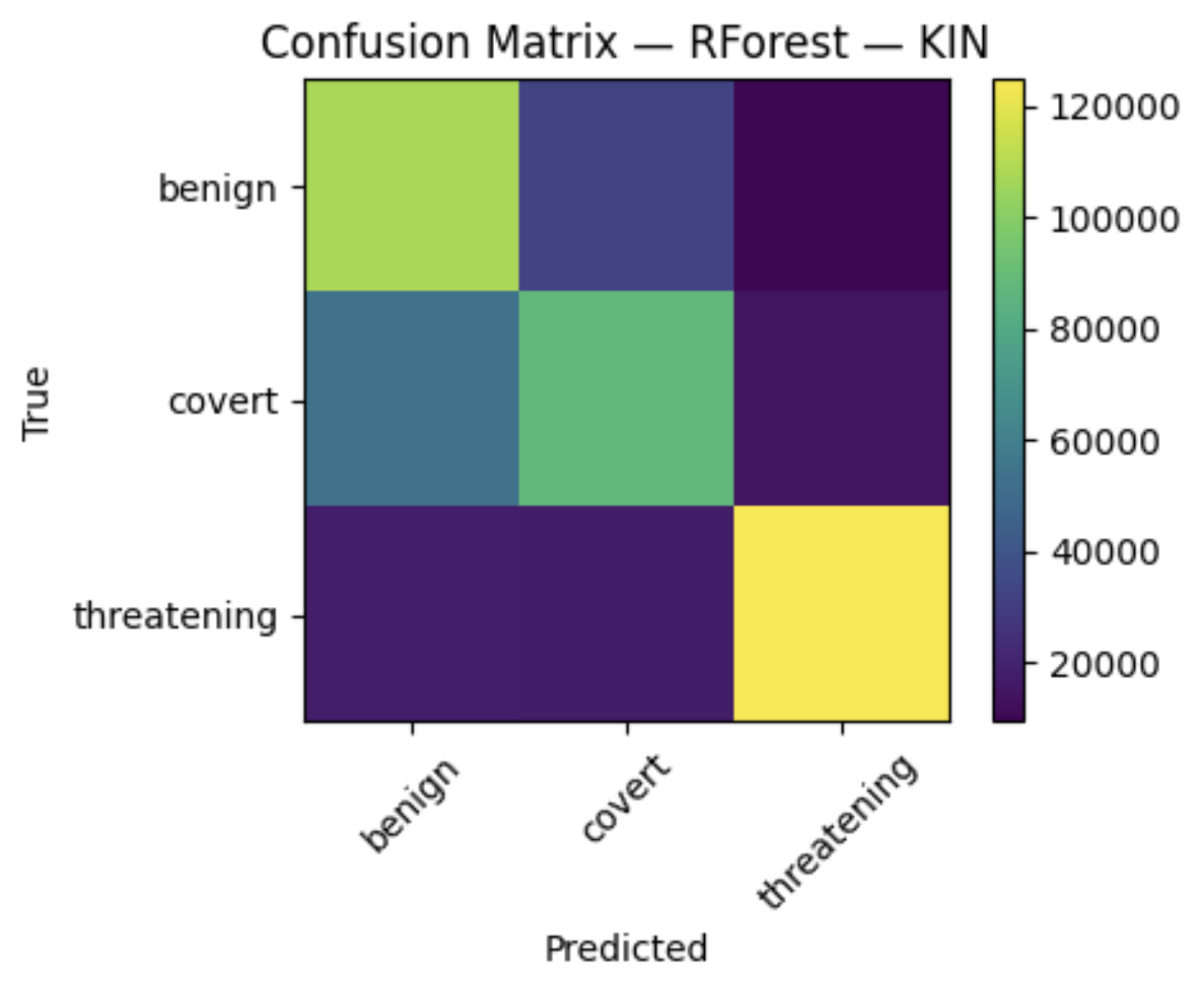}\label{fig:conf_kin}}
    \hfill
    \subfloat[Fused]{\includegraphics[width=0.33\textwidth]{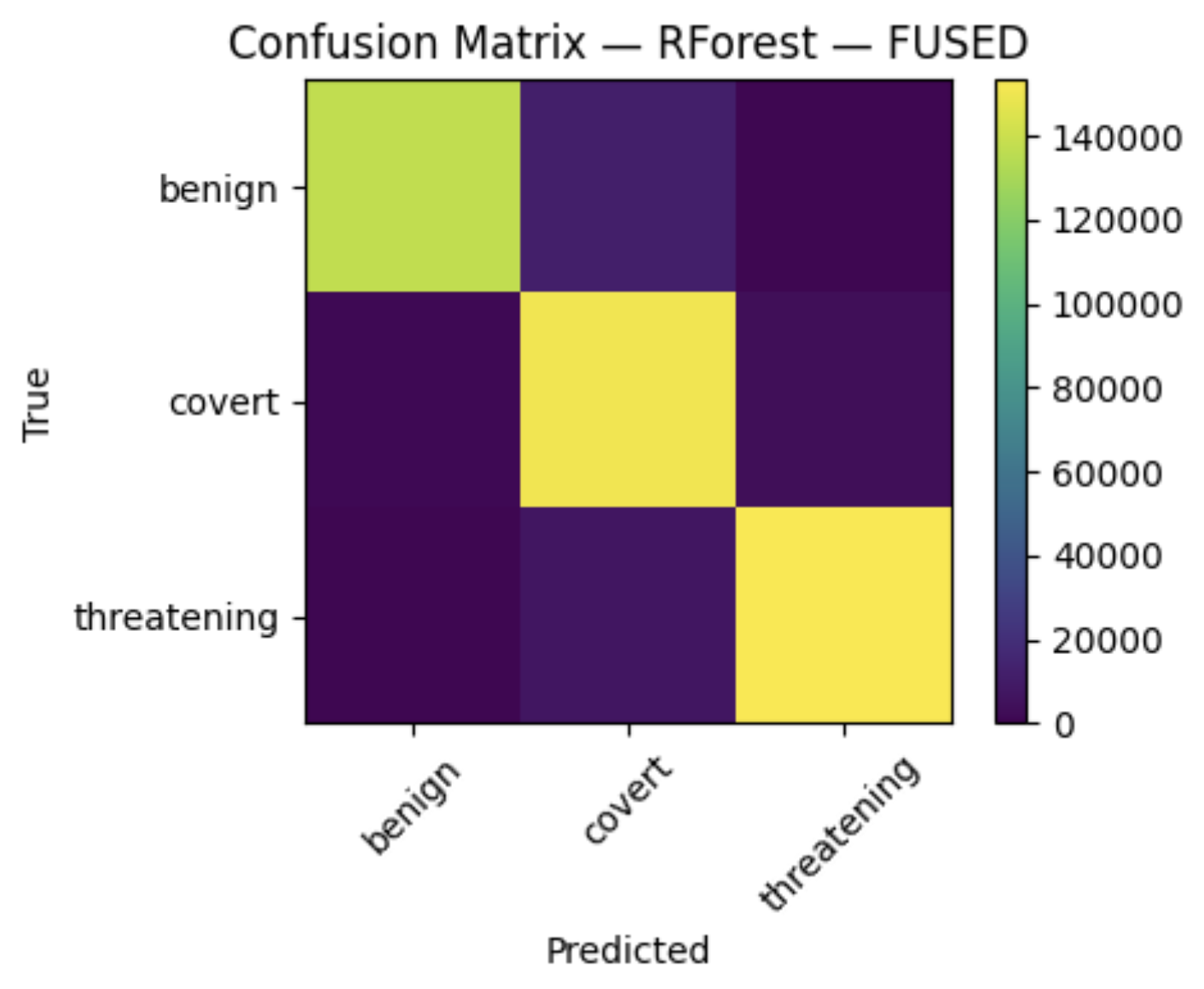}\label{fig:conf_fused}}
    \caption{(a) Confusion matrix for classification using RF-only features (RSSI, Throughput, etc.), (b) Confusion matrix using kinematic-only features (range, velocity, etc.), and (c) Confusion matrix using fused RF and kinematic features.}
    \label{fig:confusion}
\end{figure*}

To formalize the threat detection task, we define a multi-class classification function $f(\cdot)$ trained on features $\mathbf{x}_{\text{kin}}$ (kinematic) and $\mathbf{x}_{\text{RF}}$ (RF). The predicted behavior class is given by:
\begin{equation}
    b = f(\mathbf{x}_{\text{kin}}, \mathbf{x}_{\text{RF}}) \in \{\text{Benign}, \text{Covert}, \text{Threatening}\}.
    \label{eq:classifier}
\end{equation}

This formulation captures the joint decision boundary learned by the model across domains. We implement this using supervised learning models trained on labeled simulation scenarios (see Section~\ref{sec:methodology}), evaluating how well $f$ discriminates between proximity behaviors of varying threat levels. To assess the contribution of each modality, we compare model variants using RF-only, kinematic-only, and fused features. All results are reported on a held-out test set.

\subsection{Classification Performance}

We compare three model configurations:
\begin{itemize}
    \item \textbf{RF:} using only RF-derived features,
    \item \textbf{KIN:} using only kinematic features,
    \item \textbf{FUSED:} concatenating both feature sets.
\end{itemize}

As shown in Table~\ref{tab:rf_performance}, RF–RF model achieves a macro F1 score of 0.6729 and an overall accuracy of 67.56\%, with the highest precision and recall observed in the threatening class (F1 = 0.85). However, this model struggles with the covert class, exhibiting a reduced recall of 0.47, indicating difficulty in capturing low-signature, stealthy interference patterns through RF metrics alone. The RF–KIN model improves marginally, reaching a macro F1 of 0.6861 and accuracy of 68.68\%. This configuration shows better balance between classes, particularly enhancing recall for covert behaviors (from 0.47 to 0.56), suggesting that maneuver signatures are more salient for detecting subtle pursuit or shadowing behaviors. \\

Notably, the threatening class is consistently well detected across all models even in RF-only and kinematic-only configurations as seen on Fig.\ref{fig:conf_rf} and Fig.\ref{fig:conf_kin}. This likely stems from the strong signal and maneuver signatures associated with threatening behaviors—such as sustained jamming, high JSR, aggressive proximity, and elevated dynamic features like acceleration and curvature—that are easier for the model to distinguish. While fusion still improves performance significantly (e.g., F1 $>$ 0.92), the threatening class stands out as inherently more separable than covert or benign classes, which require multi-modal context to resolve effectively. Significantly, the RF–FUSED model, which incorporates both domains, achieves a macro F1 of 0.9471 and an overall accuracy of 94.67\%, outperforming all baselines by a wide margin. The F1 scores for each class exceed 0.92, confirming that the integration of temporal kinematic features and signal-level degradations provides a more discriminative representation of threat behavior. Notably, the covert class—which is traditionally the most difficult to detect—achieves a precision of 0.89 and recall of 0.96, underscoring the benefits of feature-level fusion.

\begin{table}[h]
\centering
\caption{Random Forest Performance Across Test Sets}
\label{tab:rf_performance}
\begin{tabular}{@{}lccc@{}}
\toprule
\textbf{Model} & \textbf{Accuracy} & \textbf{Macro F1} & \textbf{AUROC} \\
\midrule
RF     & 0.6756 & 0.6729 & 0.8734 \\
KIN    & 0.6868 & 0.6861 & 0.8804 \\
FUSED  & 0.9467 & 0.9471 & 0.9916 \\
\bottomrule
\end{tabular}
\end{table}

\begin{figure*}[!t]
    \centering
    \subfloat[RF-only]{\includegraphics[width=0.32\textwidth]{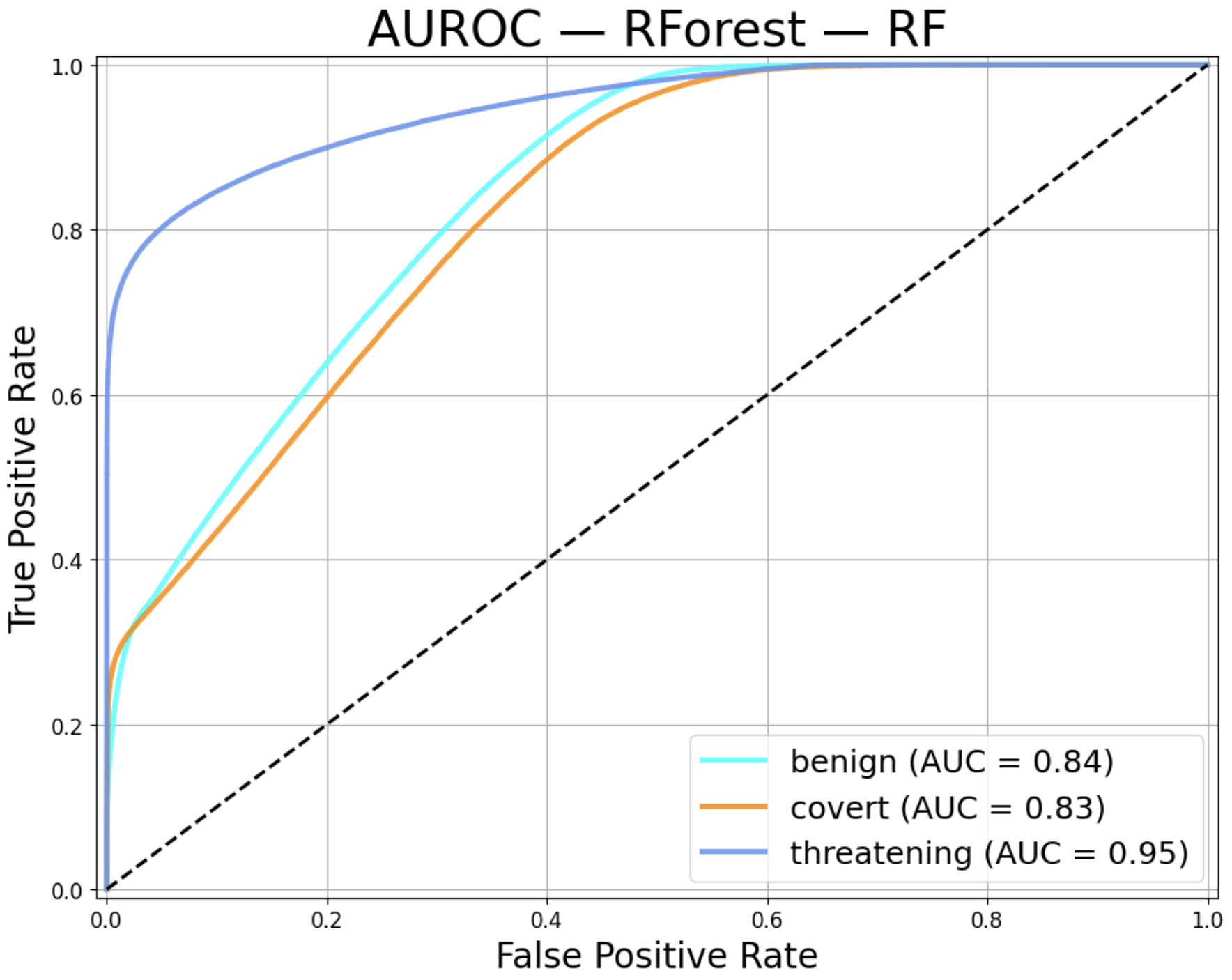}\label{fig:roc_rf}}
    \hfill
    \subfloat[Kinematic-only]{\includegraphics[width=0.32\textwidth]{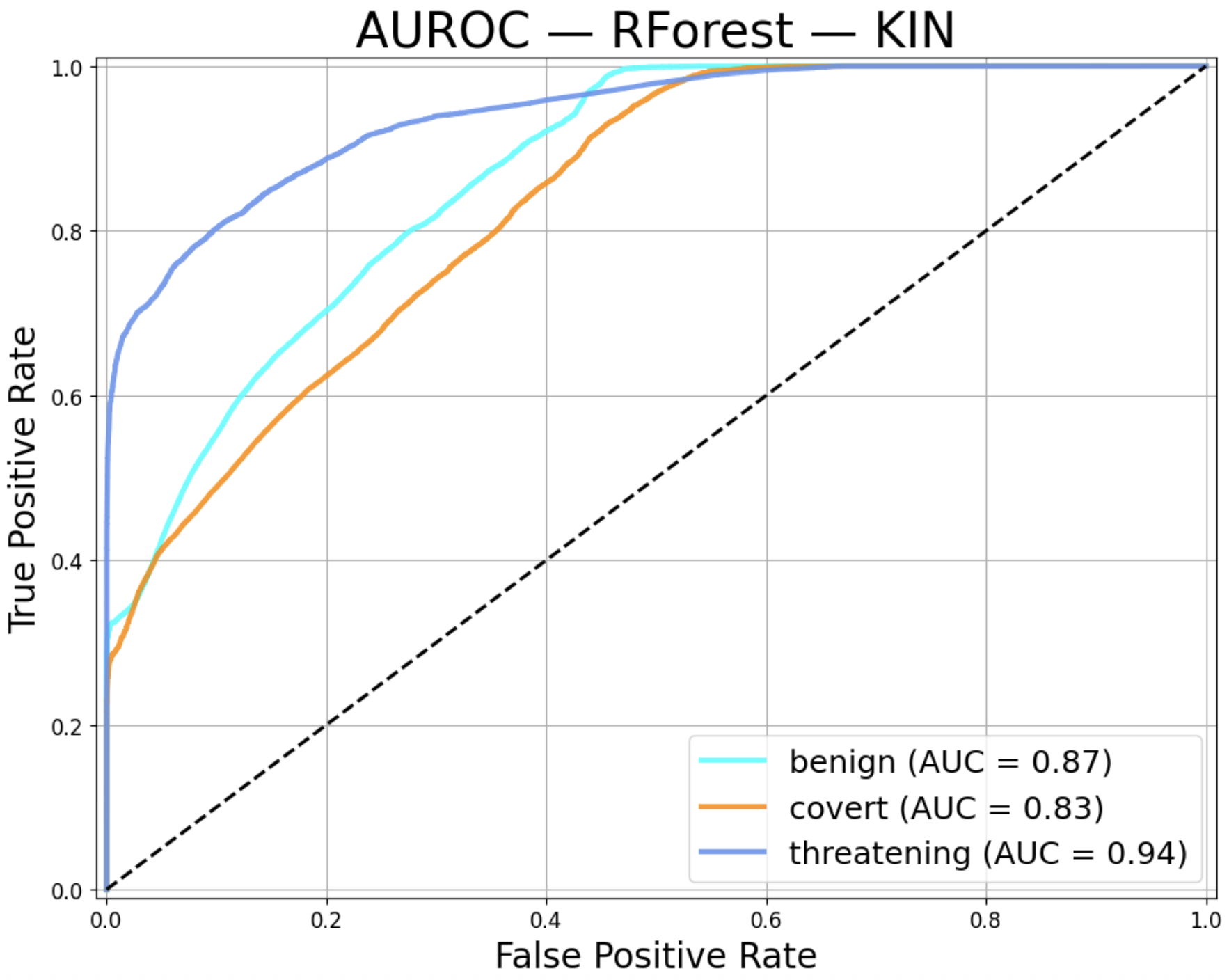}\label{fig:roc_kin}}
    \hfill
    \subfloat[Fused]{\includegraphics[width=0.32\textwidth]{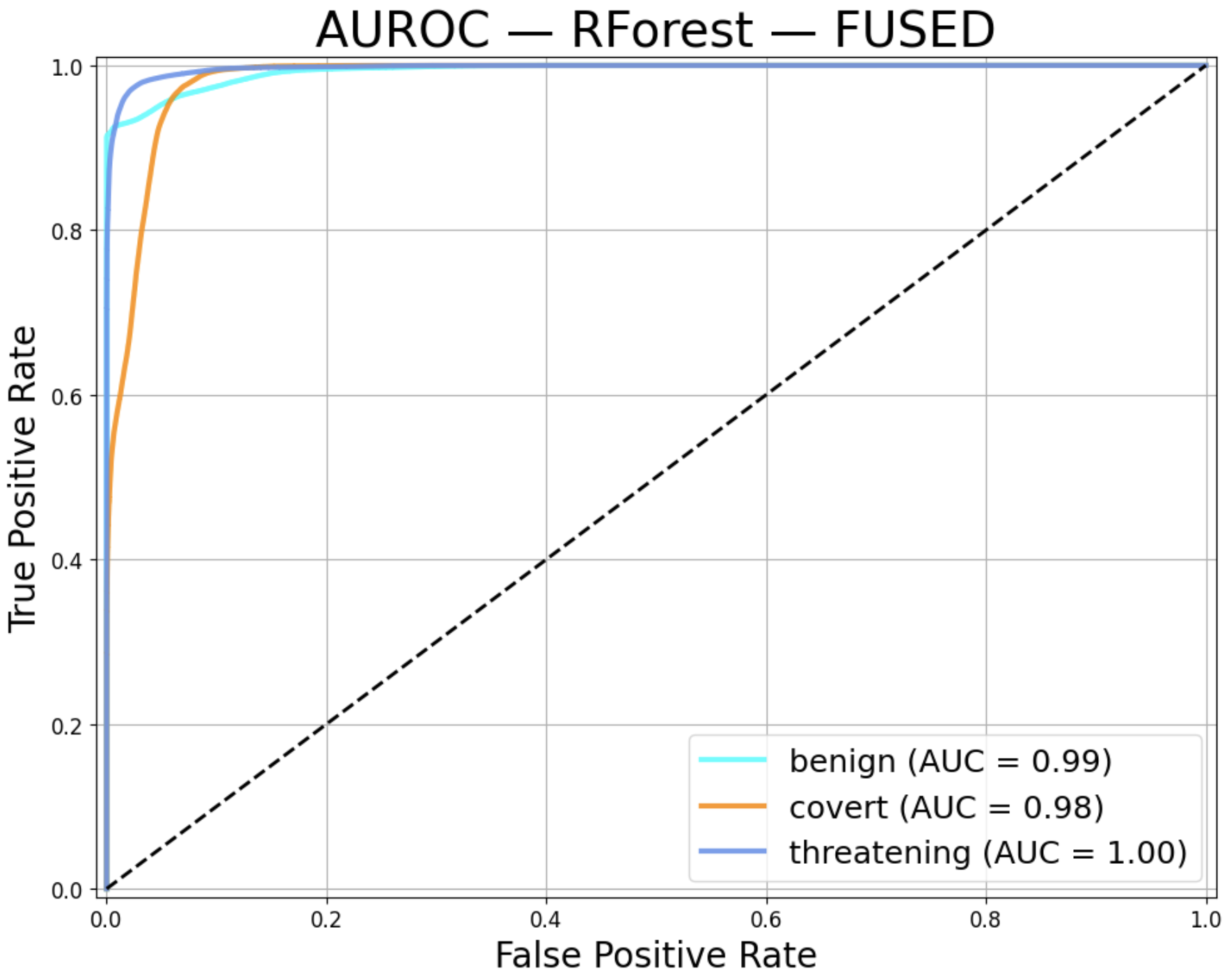}\label{fig:roc_fused}}
    \caption{(a) ROC curve for classification using RF-only features, (b) ROC curve using kinematic-only features, and (c) ROC curve using fused RF + kinematic feature.}
    \label{fig:roc}
\end{figure*}

\subsection{Class-Specific Observations (Ablation)}

In the benign class, the RF-only model tends to misclassify several samples as covert due to naturally occurring signal fluctuations; however, this confusion is significantly reduced when kinematic features are incorporated. The covert class sees the greatest improvement from data fusion, as temporal Doppler profiles and relative range rate dynamics help differentiate brief, intermittent maneuvers from background variability. For the threatening class, all models perform relatively well, but those leveraging both RF and kinematic data more effectively capture the sustained aggressiveness in both motion and interference patterns, leading to improved classification consistency.

\subsection{ROC Analysis}

Area Under the Receiver Operating Characteristic Curve (AUROC) as seen on Fig. 3 further confirms the improvement across models. The RF–FUSED model achieves a macro-averaged AUROC of 0.9916, approaching perfect separability. RF-only and kinematic-only models show similar AUROC values (0.8734 and 0.8804, respectively), indicating that each domain provides complementary yet incomplete information in isolation. The ROC curves for the kinematic-only model on Fig.\ref{fig:roc_kin} appear less smooth and exhibit more abrupt transitions than those of the RF-only model on Fig.\ref{fig:roc_rf}. This could be due to coarser or less confidently distributed output scores from the model, which may arise from noisier or more ambiguous class boundaries in the kinematic domain. Unlike RF features—where sustained jamming or signal anomalies provide strong and direct cues—kinematic indicators such as relative motion or curvature may be more subtle and temporally variable, especially in covert or benign scenarios.

\subsection{Variations Study}

In practical RF sensing systems, estimation of channel or environmental parameters at the receiver is never perfect. These imperfections, particularly those affecting received signal strength, directionality, and frequency offset, can manifest as noise or jitter in the observed features used for detection. 

While prior sections assumed reliable feature extraction, we now examine how estimation variance at the receiver affects detection performance. Crucially, this study assumes no variation in transmit power (\texttt{power\_jitter\_dB} = 0.0), to isolate the impact of receiver-side estimation errors. Instead, we sweep a set of noise scales $\sigma$ applied to key feature estimation parameters, keeping their proportional relationships intact. The variance of these estimations directly impacts the perceived feature quality.\\

The central hypothesis is that reducing the variance of estimation errors leads to significantly improved classification accuracy, particularly for low-observability classes like covert or low-jitter benign scenarios. In the limit, as $\sigma^2 \rightarrow 0$, we approach perfect sensing, and consequently, perfect detection. 

To test this, we simulate a set of increasingly noisy receiver conditions using the provided framework, applying the same Random Forest pipeline used in prior experiments. For each $\sigma$, we record key metrics on a timestep-level detection task.

\begin{figure}[h]
    \centering
    \includegraphics[width=0.45\textwidth]{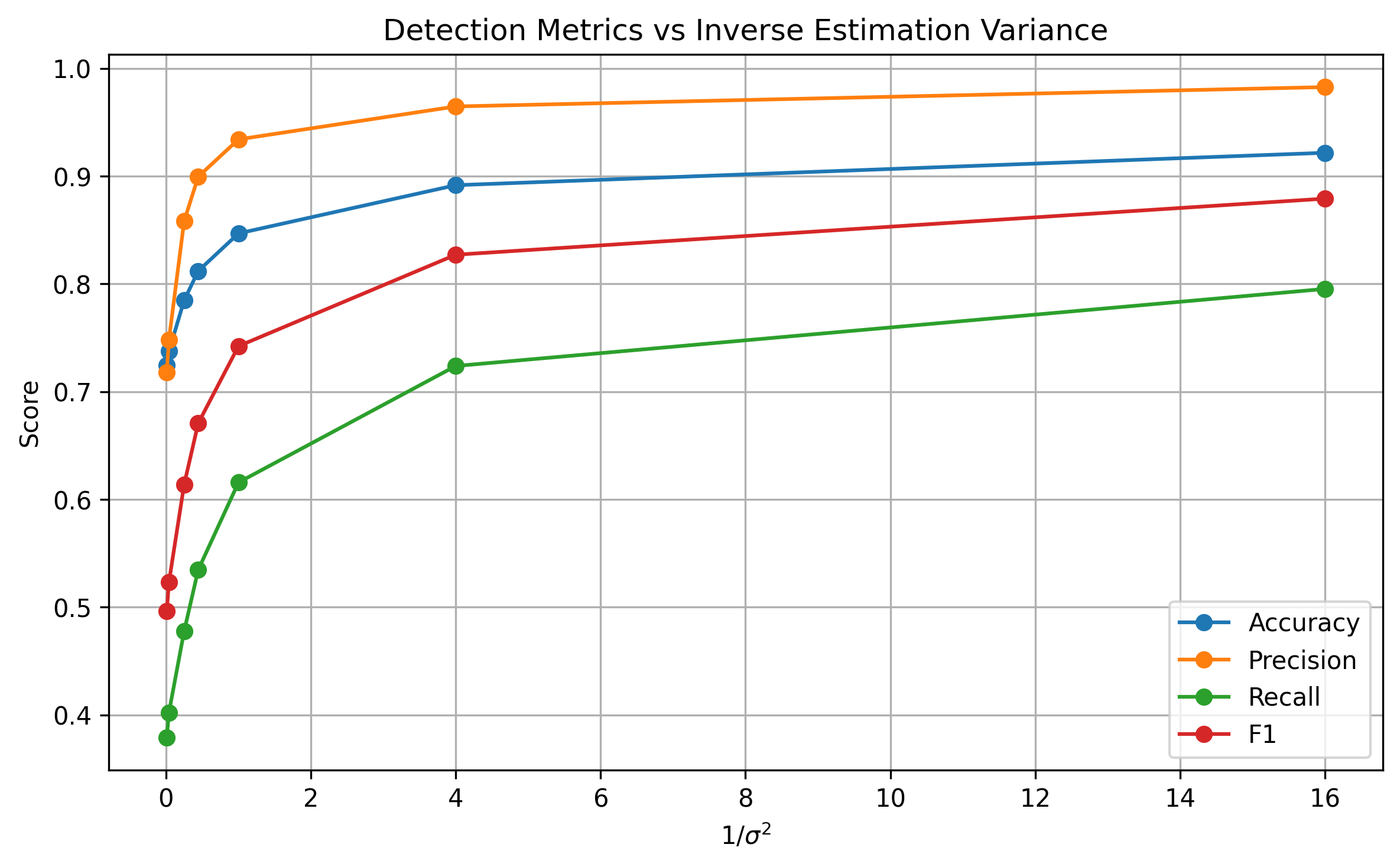}
    \caption{Detection metrics vs. inverse estimation noise variance ($1/\sigma^2$).}
    \label{fig:noise}
\end{figure}

Results confirm our expectations: when $\sigma = 0$ (i.e., no estimation error), the detection accuracy approaches 100\% (1.0 in the Fig.\ref{fig:noise} score), confirming that the signal and motion cues are intrinsically sufficient to perfectly separate jammer activity in this domain. 

As $\sigma^2$ increases, accuracy, precision, recall, and F1 all degrade predictably. Detection performance begins to fall off sharply beyond $\sigma^2 \approx 1.0$. The degradation is nonlinear, particularly for recall, which reflects a rising tendency to miss subtle, low-signature jamming behavior under noisy estimation conditions.

This trend is summarized in Fig.~4, which plots detection metrics (accuracy, precision, recall, and F1 score) as functions of the inverse estimation variance ($1/\sigma^2$). The results show that as estimation noise decreases (i.e., larger $1/\sigma^2$ values), all metrics improve consistently, approaching perfect classification when $\sigma = 0$.

The approximately linear increase of overall accuracy in the transformed space underscores the direct link between cleaner estimation and more reliable detection. These results validate the idea that channel capacity alone isn't the sole limiting factor in detection performance—rather, the fidelity of estimation plays a central role. Even when the underlying signals are strong, a noisy receiver-side view of them introduces enough ambiguity to challenge reliable classification. 

Thus, in practical deployments, efforts to reduce receiver-side estimation noise (e.g., through sensor fusion, filtering, or higher-quality RF front ends) may yield significant gains in threat detection capability, even without modifying the transmitter or increasing SNR.

\section{Discussion}
\label{sec:discussion}

The results demonstrate that joint-domain modeling provides a significant performance advantage over approaches restricted to either orbital kinematics or RF features alone. While RF-only models capture gross signal degradations associated with overt jamming, they fail to reliably identify covert interference behaviors due to their subtle temporal profiles. Conversely, kinematic-only models improve detection of low-signature behaviors but lack the ability to attribute observed dynamics to direct link-layer impacts. The fused feature set overcomes these limitations by leveraging temporal consistency between maneuver signatures and RF degradations, allowing better detection across all behavior classes. 
Importantly, the sensitivity analysis on estimation noise highlights the dependence of detection fidelity on receiver-side feature quality, emphasizing the value of robust sensor fusion and filtering. Together, these findings support the case for cross-domain fusion as a practical and scalable strategy for proximity-based interference detection.\\

Beyond simulation, several practical deployment considerations arise. First, real-world receivers are subject to estimation noise, making sensor fusion and adaptive filtering critical for operational robustness. Second, onboard implementation must balance computational demands against power and memory constraints, suggesting the need for lightweight models or hierarchical pipelines where simple classifiers trigger more detailed analysis. Third, integration with existing SSA and SatCom monitoring systems requires standardized data interfaces to ensure interoperability across operators and agencies. Finally, resilience planning demands not only accurate detection but also actionable outputs, such as maneuver advisories or adaptive link reconfiguration, to support autonomous countermeasures. Complementary to detection, recent research has explored resilience-by-design frameworks that adopt system-level planning approaches, such as the PACE methodology~\cite{boumeftah2025_pace}, to guide defensive maneuver selection and escalation strategies in proximity threat scenarios.

\section{Conclusion}
\label{sec:conclusion}

This paper presented a hybrid simulation-based framework for detecting and classifying proximity-based interference in satellite communication systems. By integrating maneuver simulation with RF link degradation modeling, we generated a comprehensive dataset of multi-orbit threat scenarios and demonstrated the benefits of feature-level fusion. Experimental results show that combining orbital kinematics with RF observables yields a substantial improvement in classification performance, particularly for covert threats that are challenging to detect in isolation. Furthermore, our analysis of estimation noise underscores the importance of high-quality receiver-side sensing for reliable detection in real-world systems. These contributions lay the groundwork for autonomous, geometry-aware interference monitoring that can enhance early warning, resilience planning, and onboard countermeasures in contested orbital environments.

\bibliographystyle{IEEEtran}
\bibliography{refs}
\nocite{*}
\end{document}